%% file: acl_latex.tex
\pdfoutput=1

\documentclass[11pt]{article}
\usepackage[final]{acl}
\usepackage{adjustbox}
\usepackage{times}
\usepackage{latexsym}
\usepackage[utf8]{inputenc}
\usepackage{microtype}
\usepackage{inconsolata}
\usepackage{graphicx}
\usepackage{amssymb}
\usepackage{algorithm}
\usepackage{algorithmic}
\usepackage{bibentry}
\usepackage{amsfonts}
\usepackage{subfigure}
\usepackage{multirow}
\usepackage{amsmath}
\usepackage{graphicx}
\usepackage{epsfig}
\usepackage{color}
\usepackage{epstopdf}
\usepackage{rotating}
\usepackage{caption}
\usepackage{float}
\usepackage{multicol}
\usepackage{amssymb}
\usepackage{graphicx}
\usepackage{bbding}
\usepackage{algorithm}
\usepackage{algorithmic}
\usepackage{balance}
\usepackage{microtype}
\usepackage{graphicx}
\usepackage{subfigure}
\usepackage{booktabs} 
\usepackage{multirow}
\usepackage{amsmath}
\usepackage{amsmath, bm}
\usepackage{tabularx}
\usepackage[T1]{fontenc}

\usepackage{mathtools}
\usepackage{etoolbox}
\usepackage{cases}
\usepackage{enumitem}
\usepackage{xcolor}
\usepackage{subfigure}
\usepackage{xcolor}
\usepackage{color}
\newcommand{\red}[1]{\textcolor{red}{#1}}
\newcommand{\blue}[1]{\textcolor{blue}{#1}}
\definecolor{brown}{RGB}{139,64,0}

\usepackage{amssymb}

\usepackage{cases}

\usepackage{cite}
\usepackage{amsmath,amssymb,amsfonts}
\usepackage{algorithmic}
\usepackage{graphicx}
\usepackage{textcomp}
\usepackage{color, colortbl}

\definecolor{backred}{RGB}{255, 190, 190}
\definecolor{backblue}{RGB}{210, 230, 250}
\definecolor{verylightgray}{gray}{0.95}

\title{Knowledge Graph Retrieval-Augmented Generation for
LLM-based Recommendation}

\author{
 \textbf{Shijie Wang\textsuperscript{1}},
 \textbf{Wenqi Fan\textsuperscript{1}}\thanks{Corresponding authors},
 \textbf{Yue Feng\textsuperscript{2}}\footnotemark[1],
 \textbf{Shanru Lin\textsuperscript{3}},
\\
 \textbf{Xinyu Ma\textsuperscript{4}},
 \textbf{Shuaiqiang Wang\textsuperscript{4}},
 \textbf{Dawei Yin\textsuperscript{4}},
\\
 \textsuperscript{1}The Hong Kong Polytechnic University,
 \textsuperscript{2}University of Birmingham,\\
 \textsuperscript{3}City University of Hong Kong,
 \textsuperscript{4}Baidu Inc,
\\
 \small{
  shijie.wang@connect.polyu.hk; wenqifan03@gmail.com; y.feng.6@bham.ac.uk; lllam32316@gmail.com; }
   \\
   \small{
  xinyuma2016@gmail.com; shqiang.wang@gmail.com; yindawei@acm.org
 }
}

\begin{document}
\maketitle
\input{sections/abs}

\input{sections/introduction-V2}
\input{sections/relatedwork}
\input{sections/model-V2}
\input{sections/experiments-V2}
\input{sections/conclusion}

\bibliography{acl_latex}

\appendix

\input{sections/appendix}

\end{document}

%% file: sections/abs.tex
\begin{abstract}
Recommender systems have become increasingly vital in our daily lives, helping to alleviate the problem of information overload across various user-oriented online services. The emergence of Large Language Models (LLMs) has yielded remarkable achievements, demonstrating their potential for the development of next-generation recommender systems. Despite these advancements, LLM-based recommender systems face inherent limitations stemming from their LLM backbones, particularly issues of hallucinations and the lack of up-to-date and domain-specific knowledge.  
Recently, Retrieval-Augmented Generation (RAG) has garnered significant attention for addressing these limitations by leveraging external knowledge sources to enhance the understanding and generation of LLMs. 
However, vanilla RAG methods often introduce noise and neglect structural relationships in knowledge, limiting their effectiveness in LLM-based recommendations.
To address these limitations, we propose to retrieve high-quality and up-to-date structure information from the knowledge graph (KG) to augment recommendations.
Specifically, our approach develops a retrieval-augmented framework, termed \textbf{K-RagRec}, that facilitates the recommendation generation process by incorporating structure information from the external KG. Extensive experiments have been conducted to demonstrate the effectiveness of our proposed method.

\end{abstract}

%% file: sections/introduction-V2.tex
 \section{Introduction}

\begin{figure}[t]
\centering
\centering
{\includegraphics[width=0.98\linewidth]{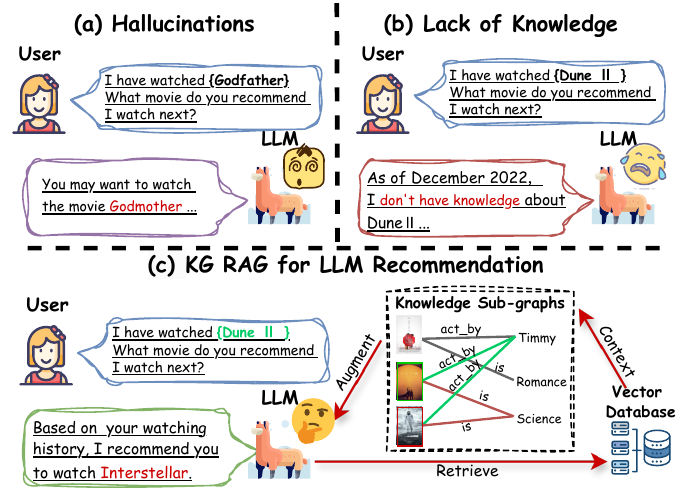}}
%
\caption{
Illustration of the issues of hallucinations and lack of domain-specific knowledge in LLM-based recommender systems and how they can be addressed by knowledge graph retrieval-augmented generation (KG RAG). } 
\label{fig:illustration}
\vskip -0.2in
\end{figure}
 
Recommender systems, as techniques designed to assist people in making decisions in their daily lives, are increasingly gaining impact in various fields~\citep{kenthapadi2017personalized,he2020lightgcn,fan2019graph}, such as online shopping, job matching, and social media.
Recently, Large Language Models (LLMs) have achieved significant breakthroughs, which further drive developments in various domains~\citep{fan2024graph,zhao2024recommender,wu2023next}. Especially with the success of LLMs, recommender systems have seen rapid growth~\citep{geng2022recommendation,bao2023tallrec,qu2024tokenrec}. By training on a wide range of data, LLMs (e.g., GPT-4~\citep{achiam2023gpt} and LLaMA~\citep{touvron2023llama}) are able to acquire extensive knowledge and demonstrate exceptional language understanding capability.
This capability enables LLM-based recommender systems to capture user preferences through a nuanced understanding of relevant attributes (e.g., user profiles, item descriptions, historical interactions) for more accurate recommendations. 
As a result, LLM-based recommender systems have emerged as a new paradigm in recommendation technology~\citep{zhao2024recommender}.

However, despite their powerful language understanding and generalization capability, LLM-based recommender systems face significant challenges, including hallucinations and lack of up-to-date and domain-specific knowledge~\citep{luo2023reasoning}. 
Specifically, one key issue is that LLM-based recommender systems may generate recommendations that are entirely fictional due to the inherent limitations of LLMs.
For example, an LLM-based recommender system may recommend \textit{"Godmother"}, a non-existent film, to the user who has watched \textit{"The Godfather"}, as illustrated in Figure~\ref{fig:illustration} (a). 
Additionally, LLMs usually lack up-to-date knowledge, which can prevent them from recommending the latest films or products in a timely manner. 
As illustrated in Figure~\ref{fig:illustration} (b), the LLM-based recommender system is unable to recommend the latest films due to the training data only containing up to December 2022. 
Furthermore, LLMs often lack domain-specific knowledge, as recommendation-oriented corpora are very limited during the training phase of LLMs~\citep{geng2022recommendation}. 
Consequently, LLMs may struggle to meet the nuanced needs of recommendation tasks. 
To alleviate these issues, one potential solution is to frequently fine-tune the LLMs with up-to-date and domain-specific knowledge. However, the massive parameters of LLMs make this process computationally expensive and time-consuming, which severely hinders the practical application in the real world.

More recently, Retrieval-Augmented Generation (\textbf{RAG}) leveraging external databases to provide specific knowledge shows promise to solve these problems~\citep{fan2024survey,gao2023retrieval}.
By incorporating an external knowledge base, RAG can retrieve relevant and up-to-date information to complement the LLM's inherent knowledge, thereby mitigating the issues of hallucinations and knowledge gaps~\citep{khandelwal2019generalization,min2020ambigqa,li2024empowering}. This makes RAG a promising technique for enabling LLMs to provide effective recommendations without the need for costly fine-tuning~\citep{ram2023context}.

Despite this potential, vanilla RAG methods that rely on documents and paragraphs often introduce unnecessary noise and even harmful disturbance, which can negatively impact the accuracy and reliability of recommendations~\citep{he2024g}. 
In addition, the structural relationships between entities are overlooked in typical RAG, resulting in the sub-optimal reasoning capability of LLM-based recommender systems~\citep{luo2023reasoning}.
To address the limitations, a prospective solution is to incorporate structured knowledge such as \emph{items' knowledge graph (KG)} to help improve recommendation performance. 
Specifically, KGs offer structured, factual, and editable representations of knowledge, which can provide a faithful knowledge source for recommendations.
As shown in Figure~\ref{fig:illustration} (c), retrieving structured knowledge from the KG can significantly enhance the recommendation capabilities of LLM-based recommender systems.

However, it is challenging to effectively retrieve KGs to enhance the recommendation capabilities of LLMs.
First, KGs store rich factual knowledge in a structured format.
Simply retrieving the triplets or first-order neighbors of an entity (i.e., item) with semantic information neglects the importance of these higher-order neighborhood effects among entities/items, resulting in sub-optimal recommendation performance.
Second, indiscriminate retrieving for each item, regardless of whether the retrieval is necessary or the content is relevant, can degrade the performance of the recommendation while severely reducing the model's efficiency~\citep{asai2023self,labruna2024retrieve}.
Furthermore, structured data in KGs is typically encoded for LLM in the form of serialized text~\citep{wu2023retrieve,sun2023think}, which is insufficient for fully exploiting the structural information inherent in the data~\citep{perozzi2024let,fatemi2023talk}.
Therefore, it is crucial to explore more effective and expressive ways of representing structured data, allowing LLMs to effectively leverage the structure information of retrieved knowledge sub-graphs for recommendations.

To address the aforementioned challenges, in this paper, we propose a knowledge retrieval-augmented recommendation framework, namely \textbf{K-RagRec}, to provide up-to-date and reliable knowledge by retrieving relevant knowledge from item's KGs for recommendation generation in the era of LLMs. 
Specifically, our proposed framework first performs knowledge sub-graph indexing on the items' KG at a coarse and fine granularity to construct the knowledge vector database. 
Next, a popularity selective retrieval policy is designed to determine which items should be retrieved, followed by the retrieval of specific sub-graphs from the knowledge vector database. 
To refine the quality of retrieval and ensure the most relevant results are prioritized at the top of the input, we subsequently re-rank the retrieved knowledge sub-graphs. 
Finally, we introduce a GNN and projector to align the retrieved knowledge sub-graphs into the semantic space of the LLM for knowledge-augmented recommendation.
The main contributions of this paper can be summarized as follows:
\vspace{-0.1in}
 \begin{itemize} [leftmargin=*] 
\setlength{\itemsep}{0pt}
\setlength{\parsep}{0pt}
\setlength{\parskip}{0pt}
     \item We propose a novel framework that retrieves faithful knowledge from KGs to augment the recommendation capability of LLM. 
     Note that we introduce a flexible indexing method for KGs, which can provide a comprehensive view of a node’s neighborhood within KG.
     \item We design a popularity selective retrieval strategy to determine whether an item needs to be retrieved based on its popularity, significantly improving efficiency.
     \item We introduce a more expressive graph encoder for structured data inclusion in LLMs, that can facilitate the LLM to effectively leverage the structure information and avoid long context input.
     \item We conduct comprehensive experiments on various real-world datasets to evaluate the effectiveness of the proposed K-RagRec framework.
\end{itemize}

%% file: sections/relatedwork.tex
\section{Related Work} 

Recently, RAG has emerged as one of the most representative technologies in the field of generative AI, combining the strengths of retrieval systems and language models (LM) to produce coherent and informative text.
Early methods, such as REALM~\citep{guu2020retrieval},  RETRO~\citep{borgeaud2022improving}, and DPR~\citep{karpukhin2020dense}, typically involve retrieving relevant fragments from a large corpus to guide the LM generation. 
However, standard RAG methods often struggle to accurately retrieve all relevant textual chunks, due to unnecessary noise and even harmful disturbance in the documents.
To address these limitations,  recent studies~\citep{baek2023knowledge,wu2023retrieve,he2024g,luo2023reasoning,wang2023knowledge,sen2023knowledge,sun2023think} focus on retrieving structured and faithful knowledge from graphs for enhancing generations. For example, Retrieve-Rewrite-Answer~\citep{wu2023retrieve} retrieves relevant sub-graphs from KG and converts retrieved sub-graphs into text for the generation.
G-Retriever~\citep{he2024g} explores retrieving sub-graphs from various types of graphs to alleviate the hallucinations of LLM.

With the explosion of LLMs in recommendations, a few works~\citep{di2023retrieval,wu2024coral,ang2024tsgassist} make initial explorations in RAG for recommendations.
For instance, the work~\citep{di2023retrieval} proposes
leveraging knowledge from movie or book datasets to enhance recommendations. 
Nevertheless, retrieving faithful structured knowledge from the KGs for recommendations is under-explored and shows great potential.
To fill this gap, we propose to retrieve knowledge sub-graphs from KGs to enhance the recommendation performance of LLM. We provide more comprehensive related work on Appendix~\ref{appendix:related_work}.

%% file: sections/model-V2.tex
\section{Methodology}
In this section, we first introduce some key notations and concepts in this work.
Then, we provide the details for each component of our proposed framework K-RagRec.

\subsection{Preliminary}

\textbf{Knowledge Graph:}
In this work, we propose to leverage external knowledge databases (i.e., KGs) to augment LLMs for recommendations.
KGs contain abundant factual knowledge in the form of a set of triples: $\mathcal{G}=\{(n,e,n') \ | \ n,n' \in N \ , \ e \in E\}$, where $N$ and $E$ denote the set of entities and relations, respectively. For example, a triple $Interstellar \ \xrightarrow{directed\_by} \ Nolan$ indicates that the movie \textit{"Interstellar"} was directed by the director \textit{"Nolan"}.

\noindent \textbf{LLM-based Recommendations:}
Let $U=\{u_1,u_2,\dots,u_n\}$ and $V=\{v_1,v_2,\dots,v_m\}$ represent the sets of users and items, where $n$ and $m$ are the sizes of users and items, respectively. 
The goal of an LLM-based recommender system is to understand users' preferences by modeling users' historical items interactions $V^{u_i}=[v^{u_i}_1, v^{u_i}_2, \cdots,v^{u_i}_{|L_{u_i}|} ]$ (\textit{e.g.}, clicks, bought, etc.), where $|L_{u_i}|$ is interaction sequence length for user $u_i$.  
Notably,  item $v_j$' side information, such as title and description, is publicly available to enhance LLM for modeling user preferences.  
In our setting, we consider asking the LLM to select user $u$'s preferred item $v$ from the candidate set $C = \{v_j\}^M_{j=1}$, where $M$ is the number of candidate items. The candidate set $C$ typically consists of one positive sample as well as $M-1$ negative samples.
Specifically, for a frozen LLM $f_{\delta}$ with parameters $\delta$, we denote an input-output sequence pair as $(Q, A)$, where $Q$ is the recommendation query/prompt, which consists of task descriptions and users' historical items. The output $A$ is the LLM's prediction. Furthermore, we introduce the concept of GNNs in appendix~\ref{appendix:gnn}.

\subsection{The Overview of Proposed Method}
As shown in Figure~\ref{fig:framework}, our proposed K-RagRec consists of five crucial components: \emph{Hop-Field Knowledge Sub-graphs for Semantic Indexing, Popularity Selective Retrieval Policy, Knowledge Sub-graphs Retrieval, Knowledge Sub-graphs Re-Ranking, and Knowledge-augmented Recommendation.} The model first performs indexing of hop-field knowledge sub-graphs within the KG. 
Following this, a popularity selective retrieval policy is implemented to determine which items should be retrieved or augmented. The model then retrieves specific sub-graphs from the knowledge vector database. 
Subsequently, the retrieved knowledge sub-graphs are re-ranked to refine the retrieval quality. Finally, the retrieved knowledge sub-graphs are utilized with the original prompt to generate recommendations.

\subsection{Hop-Field Knowledge Sub-graphs for Semantic Indexing} 
\begin{figure*}[htb]
    \centering
    \vskip -0.1in
    \includegraphics[width=0.95\textwidth]{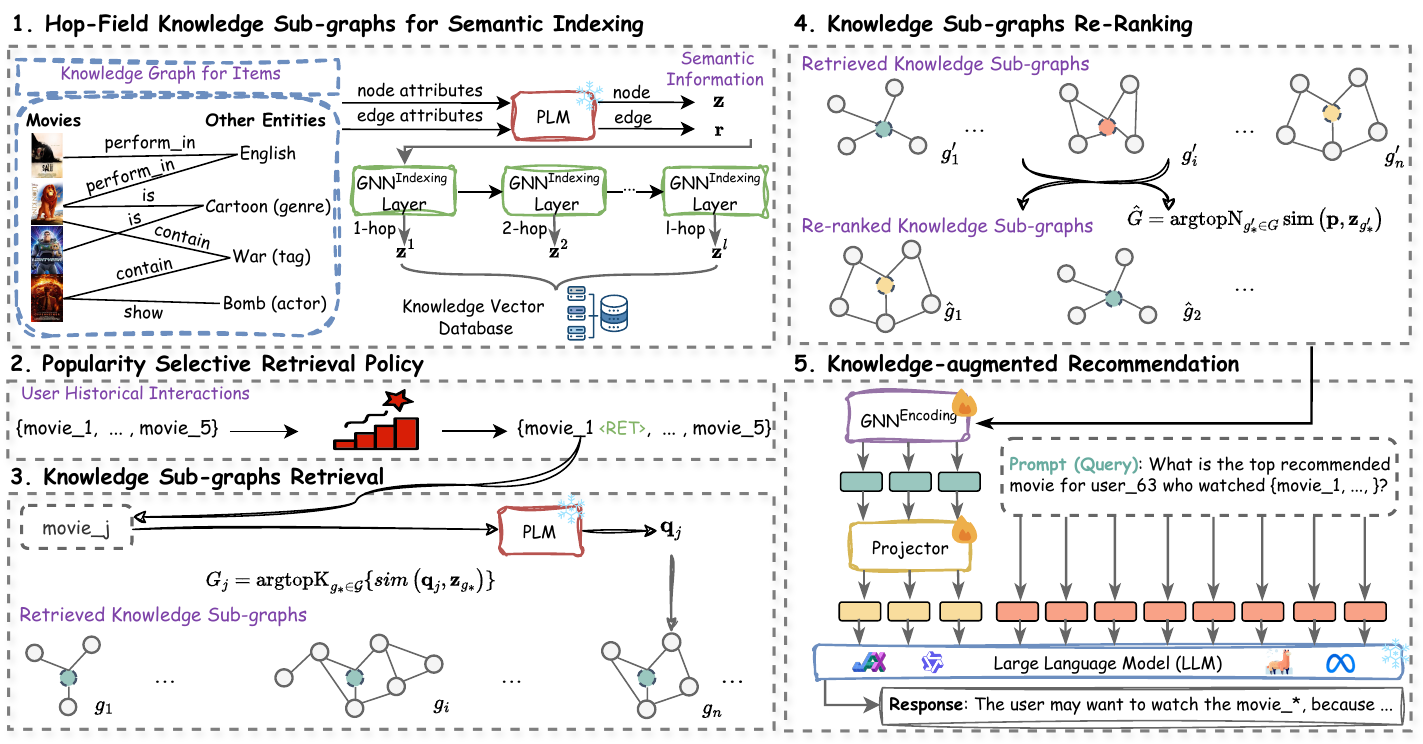}
    \vskip -0.1in
    \caption{The overview of the K-RagRec. 
It contains five key components: \emph{Hop-Field Knowledge Sub-graphs for Semantic Indexing, Popularity Selective Retrieval Policy, Knowledge Sub-graphs Retrieval, Knowledge Sub-graphs Re-Ranking, and Knowledge-augmented Recommendation.}
    }
    \label{fig:framework}
    \vskip -0.2in
\end{figure*}

Typically, retrieving knowledge from KG chunks the KG into nodes~\citep{he2024g} or triplets~\citep{luo2023reasoning} and only retrieves content locally around the target entity in the KG.
However, these methods just naively retrieve the first-order neighbors of an entity (i.e., item), which makes it difficult to capture these higher-order neighborhood effects among entities/items in the recommendation process. 
Therefore, to effectively retrieve knowledge from the KG, we propose performing semantic indexing on hop-field knowledge sub-graphs, which can flexibly chunk KGs and provide a comprehensive view of a node's neighborhood in KG.
As illustrated in Figure~\ref{fig:framework} component 1, we first introduce a pre-trained language model (PLM), such as SentenseBert~\citep{reimers2019sentence},  to capture the semantic information for node $n_o$ as follows:
\begin{equation}
\setlength{\abovedisplayskip}{3pt}
\setlength{\belowdisplayskip}{3pt}
\textbf{z}_{n_o}=\text{PLM}(x_{n_o})\in {R}^d,
\end{equation}
where $d$ is the dimension of the output representation. Similarly, we also capture the semantic information for edge/relation $e_o$ in KG:
\begin{equation}
\setlength{\abovedisplayskip}{3pt}
\setlength{\belowdisplayskip}{3pt}
\textbf{r}_{e_o}=\text{PLM}(x_{e_o})\in {R}^d,
\end{equation}
where $x_{n_o}$ and $x_{e_o}$ are the text attributes (e.g., item's title and descriptions) of node $n_o$ and edge/relation $e_o$, respectively.

To retrieve nuanced knowledge of both coarse and fine graph structures from KG, we introduce a GNN (i.e., $\text{GNN}_{\phi_1}^{\text{Indexing}} (\cdot)$ with parameters $\phi_1$) to aggregate information from neighbors for entities, where the $l_{th}$-hop embedding $\textbf{z}_{n_o}^{(l)}$ of a central entity $n_o$ can be defined by:
\begin{small}
\begin{equation}
\setlength{\abovedisplayskip}{3pt}
\setlength{\belowdisplayskip}{3pt}
\textbf{z}_{n_o}^{l} = \text{GNN}_{\phi_1}^{\text{Indexing}} (\{\textbf{z}_{n_m}^{(l-1)}, \textbf{r}_{e_{<o, m>}}^{(l-1)}: \\
~ n_m \in \mathcal{N}(n_o)   \}),
\end{equation}
\end{small}
where $\mathcal{N}(n_o)$ is the set of neighbours of node $n_o$, 
and $e_{<o, m>}$ is the edge between node $n_o$ and $n_m$.
For each entity, its $l$-hop representation can be seen as a knowledge sub-graph representation containing the $l$-hop neighbors of itself. 
Therefore, we can express the knowledge sub-graph representation of $g_o \in \mathcal{G}$ as $\textbf{z}_{g_o}$, where $\mathcal{G}$ is the set containing all the knowledge sub-graphs. For each sub-graph, we store its representation in a \emph{knowledge vector database}.

 \subsection{Popularity Selective Retrieval Policy}
Although RAG can augment the LLM for modeling user preferences with retrieved knowledge, retrieving each item can cost a significant amount of retrieval time, which can severely degrade the user experience and cause user churn.
Meanwhile, most users’ online behaviours in recommender systems are following the power law distribution~\citep{abdollahpouri2017controlling,celma2008new} in which a small proportion (e.g., less than 20\%) of items (i.e., popular items) often account for a large proportion of users’ online behaviours (e.g., more than 80\%), while cold-start items have a few interactions from users.
Therefore, most models tend to keep rich knowledge of popular items, resulting in an inferior performance for cold-start items~\citep{zhao2023popularity}. 
To this end, we design a popularity selective retrieval policy to determine whether an item needs to be augmented from KG based on its popularity (e.g., sales volume and page view). 
Particularly, the item is retrieved if its popularity is less than the pre-defined threshold $p$, otherwise not. 
By incorporating this strategy, the retrieval time in K-RagRec can be significantly reduced to achieve more efficient retrieval.

 \subsection{Knowledge Sub-graphs Retrieval} 
Given the query for knowledge sub-graphs retrieval, we adopt the same PLM as the first indexing step to ensure that the query is in the consistent embedding space as the knowledge sub-graph representation. 
We define the text attribute (e.g., item's title and descriptions) $x_{q_j}$ of the item $v_j$ that needs to be retrieved as the query and obtain its semantic information $\textbf{q}_j$ as: 
\begin{equation}
\setlength{\abovedisplayskip}{2pt}
\setlength{\belowdisplayskip}{2pt}
\textbf{q}_j=\text{PLM}(x_{q_j})\in {R}^d.
\end{equation}

Next, we retrieve the top-$K$ most similar sub-graphs $G_j=\{g'_1, ..., g'_K\}$ from  $\mathcal{G}$ for item $v_j$:  
\begin{equation}
\setlength{\abovedisplayskip}{3pt}
\setlength{\belowdisplayskip}{3pt}
G_{j} = \operatorname{argtopK}_{g_* \in \mathcal{G}} \text{sim} \left(\textbf{q}_j, \textbf{z}_{g_*} \right),
\end{equation}
where  ${\text{sim}(\cdot, \cdot)}$ is a similarity metric for measuring the similarity between the query representation $\textbf{q}_j$ and knowledge sub-graph ${g_*}$'s representation $\textbf{z}_{g_*}$  in knowledge vector database. 
Finally,  the retrieved knowledge sub-graphs for items required to be retrieved from user ${u_i}$'s historical interactions $V^{u_i}$ will be used to form a knowledge sub-graph set $G$.

\subsection{Knowledge Sub-graphs Re-Ranking} 
To refine the quality of retrieval for enhancing recommendation performance, the next crucial step is knowledge sub-graphs re-ranking.
Feeding all retrieved knowledge sub-graphs $g' \in G$ directly into LLM can lead to information overload if the user ${u_i}$ has a long historical interaction list towards items.
Therefore, we execute re-ranking to shorten the retrieved knowledge sub-graphs and ensure the most relevant knowledge sub-graphs are prioritized at the top of the prompt. 
Specifically, we adopt the recommendation prompt as a query for re-ranking, which consists of \textbf{task descriptions} and \textbf{users’ historical items}.
For example, this recommendation prompt can be \textit{"What is the top recommended movie for the user who watched \{Matrix, ..., Iron Man\}?"}. 
For consistency, we adopt the same PLM to capture the semantic information of the above prompt as $\textbf{p}$, and re-rank the knowledge sub-graphs in $G$ to obtain a  Top-$N$ knowledge sub-graphs set $\hat{G}$: 
\begin{equation}
\setlength{\abovedisplayskip}{3pt}
\setlength{\belowdisplayskip}{3pt}
\hat{G}=\operatorname{argtopN}_{g_*' \in G} \text{sim} \left(\textbf{p}, \textbf{z}_{g_*'} \right).
\end{equation}

 \subsection{Knowledge-augmented Recommendation}
To facilitate the LLM's better understanding of the structure of retrieved knowledge sub-graphs and to avoid long contexts, we further integrate another  GNN encoder $\text{GNN}^{\text{Encoding}}_{\phi_2}$ with parameter ${\phi_2}$ to enhance the representation learning of structural information:
\begin{equation}
\setlength{\abovedisplayskip}{3pt}
\setlength{\belowdisplayskip}{3pt}
\textbf{h}_{\hat{g_*}}=\text{GNN}^{\text{Encoding}}_{\phi_2}(\{\hat{g_*}: \hat{g}_*\in \hat{G}\}).
\end{equation}

An MLP projector $\text{MLP}_{\theta}$ with parameter $\theta$ is further introduced to shift to mapping all sub-graphs embedding in $\hat{G}$ into the LLM embedding space:

\begin{equation}
\setlength{\abovedisplayskip}{2pt}
\setlength{\belowdisplayskip}{2pt}
\hat{\textbf{h}}_{\hat{G}}= \text{MLP}_{\theta}([\textbf{h}_{\hat{g_1}};...;\textbf{h}_{\hat{g_N}}]),
\end{equation}
where $[.; .]$ represents the concatenation operation. 
The extracted knowledge sub-graphs embedding $\hat{\textbf{h}}_{\hat{G}}$ as the soft prompt is then appended before the input token embedding in LLM.

\subsection{Optimization for K-RagRec} 
The training process can be broadly considered as soft prompt tuning, where the retrieved knowledge sub-graphs are a series of soft graph prompts.
Formally, the generation process can be expressed as follows:
\begin{small} 
\begin{equation}
\setlength{\abovedisplayskip}{3pt}
\setlength{\belowdisplayskip}{3pt}
p_{\delta, \theta, \phi_1, \phi_2}(Y \mid \hat{G}, x_q) =
\prod_{k=1}^r p_{\delta, \theta, \phi_1, \phi_2}(y_k \mid y_{<k}, \hat{\textbf{h}}_{\hat{G}}, x_q).
\end{equation}
\end{small}

Therefore, instead of fine-tuning the LLM model extensively, we only learn parameters of two GNNs (i.e., $\text{GNN}_{\phi_1}^{\text{Indexing}}$, $\text{GNN}_{\phi_2}^{\text{Encoding}}$) and projector  $\text{MLP}_\theta$, while parameters $\delta$ of LLM backbone are frozen. We update the parameters $\phi_1$, $\phi_2$ and $\theta$ through the Cross-Entropy Loss $\mathcal{L}(Y,A)$, where $Y$ is the ground-truth and $A$ is LLM's prediction.

%% file: sections/experiments-V2.tex
\section{Experiment}
In this section, we evaluate the effectiveness of our proposed framework through comprehensive experiments. 
First, we present the experimental settings, including details about the datasets, compared baselines, evaluation metrics, and the parameter configurations. Then, we report the main experimental results, highlighting the performance of the proposed framework compared with various baseline methods.
Finally, we analyze the contributions of individual model components and the impact of parameters used in our framework. We also assess the generalizability of K-RagRec in the zero-shot setting. We present the generalizability study in Appendix~\ref{appendix:General}.

\subsection{\textbf{Experimental Settings}}
\subsubsection{\textbf{Datasets}}\label{datasets}
To evaluate the performance of our K-RagRec framework, we adopt three real-world datasets.
\textbf{MovieLens-1M}\footnote{https://grouplens.org/datasets/movielens/1m/} is a dataset containing approximately one million movie ratings and textual descriptions of movies (i.e., “title”). 
\textbf{MovieLens-20M}\footnote{https://grouplens.org/datasets/movielens/20m/} is a large-scale movie ratings dataset encompassing over 20 million ratings from more than 138,000 users on 27,000 movies.
\textbf{Amazon Book}\footnote{https://jmcauley.ucsd.edu/data/amazon/} is a book recommendation dataset that records more than 10 million user ratings of books and the titles of the books.
In addition, we adopt the popular knowledge graph \textit{Freebase}\footnote{https://developers.google.com/freebase} and filter out the triples related to the three datasets to reconstruct the KG. The statistics of these three datasets and KG are presented in Table~\ref{tab:datasets}.


\subsubsection{\textbf{Baselines}}\label{baseline}

In the realm of LLM-based recommendation research, our work pioneers the investigation of retrieving knowledge from KGs to enhance the recommendation capabilities of LLMs.
Therefore, to evaluate the effectiveness, we compare our proposed framework with a series of meticulously crafted KG RAG-enhanced LLM recommendation baselines.
We first include two typical inference-only methods Retrieve-Rewrite-Answer~\citep{wu2023retrieve} (KG-Text) and KAPING~\citep{baek2023knowledge}, where the former retrieves sub-graphs and textualizes them, and the latter retrieves triples.
We exclude some knowledge reasoning path-based approaches~\citep{luo2023reasoning}, as it is difficult to retrieve faithful knowledge reasoning paths solely from the user's interaction items.
Next, we compare K-RagRec with various prompt-tuning approaches augmented by retrieval, including Prompt Tuning with KG-Text (PT w/ KG-Text), GraphToken~\citep{perozzi2024let} with retrieval (GraphToken w/ RAG) as well as G-retriever~\citep{he2024g}. Additionally, we evaluate our method against Lora Fine-tuning with retrieval (Lora w/ KG-Text)~\citep{hu2021lora}.

\subsubsection{\textbf{Evaluation Metrics}}
To evaluate the effectiveness of our K-RagRec framework, we employ two widely used evaluation metrics: Accuracy (ACC), and Recall@$k$ ~\citep{he2020lightgcn}. We present results for $k$ equal to 3, and 5. Inspired by recent studies~\citep{zhang2024agentcf,hou2022towards}, we adopt the \textit{leave-one-out strategy} for
evaluation. Specifically, for each user, we select the last item that the user interacted with as the target item and the 10 interaction items prior to the target item as the historical interactions. Then, we leverage LLM to predict the user's preferred item from a pool of 20 candidate items ($M = 20$), which contains one target item with nineteen randomly sampled items. For trained models (including prompt tuning and fine-tuning), we compute Recall@$k$ by extracting the probability assigned to each item and evaluating the model's ability to rank the target item within the top-$k$ predictions. In addition, we also conduct comparison experiments with 10 candidate items ($M = 10$) as shown in Appendix~\ref{appendix:10samples}.

\input{tables/main_results}

\subsubsection{Parameter Settings}
We implement the framework on the basis of PyTorch and conduct the experiments on 2 NVIDIA A6000-48GB GPUs. We adopt the SentenceBert to encode entities, relations, and query attributes. 
We use the 3 layers Graph Transformer as the $\text{GNN}^{\text{Indexing}}$ and $\text{GNN}^{\text{Encoding}}$ for MovieLens-1M and 4 layers for MovieLens-20M and Amazon Book. The layer dimension is set to 1024, and the head number is set to 4. 
The popularity selective retrieval policy threshold $p$ is set to 50\%.
For each item that needs to be retrieved, we retrieve the top-3 most similar sub-graphs. The re-ranking knowledge sub-graph number $N$ is set to 5. More experiment details are shown in Appendix~\ref{appendix:hyper}. We also present
several prompt examples in Appendix~\ref{appendix:prompt}.

\subsection{\textbf{Overall Performance Comparison}}
We compare the recommendation performance of K-RagRec with various baselines on three open-source backbone LLMs: LLama-2-7b~\citep{touvron2023llama}, LLama-3-8b~\citep{dubey2024llama}, and QWEN2-7b~\citep{yang2024qwen2}. We present the results in Table~\ref{tab:results}. From the comparison, we have the following main observations:
\vspace{-0.1in}
 \begin{itemize} [leftmargin=*] 
\setlength{\itemsep}{0pt}
\setlength{\parsep}{0pt}
\setlength{\parskip}{0pt}
\item Naively retrieve KG and augment LLM with text methods (i.e., KG-Text and KAPING),  have limited recommendation accuracy on the MovieLens-1M and MovieLens-20M and Amazon Book datasets.

\item Compared to other prompt tuning RAG methods, K-RagRec with LLama-2-7B as the backbone LLM leads to an average of 41.6\% improvement over the sub-optimal baseline across all datasets. With LLama-3-8B and QWEN2-7B as the backbone LLM, K-RagRec also brought an average of 13\% to 32\% improvement, highlighting the effectiveness of our proposed method in augmenting LLMs' recommendation performance.

\item Compared to the LoRA fine-tuning with the naive RAG approach, K-RagRec with prompt tuning achieves close to or better performance in most settings. Notably, K-RagRec achieves the best performance when fine-tuned with LoRA.
\end{itemize}


\begin{figure}[t]
\vskip -0.1in
\centering
\subfigure[ML1M ACC]{\includegraphics[width=0.32\columnwidth]{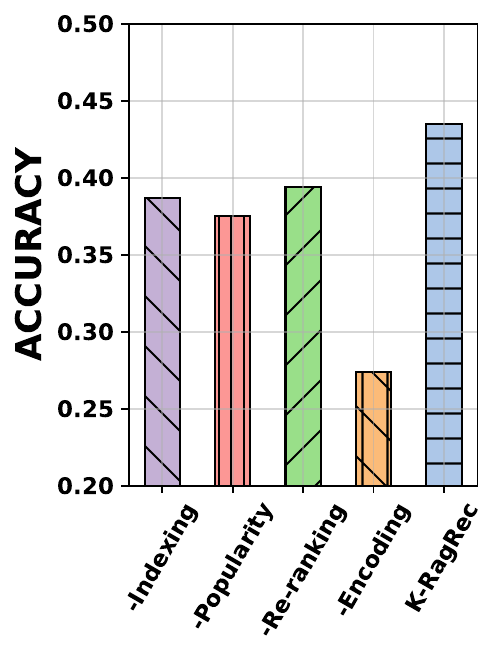}}
\subfigure[ML1M R@3]{\includegraphics[width=0.32\columnwidth]{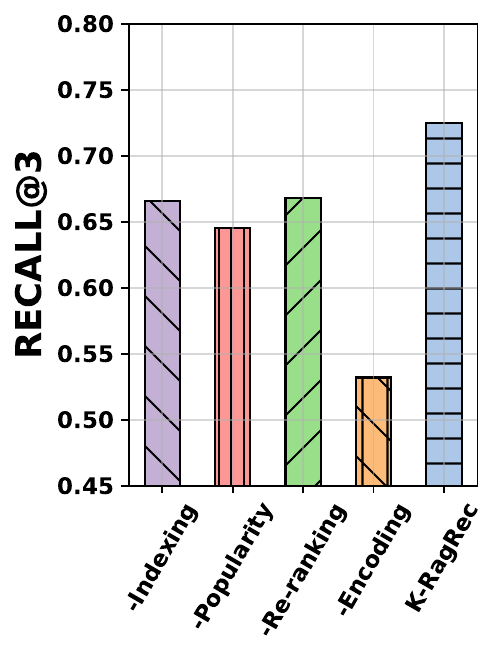}}
\subfigure[ML1M R@5]{\includegraphics[width=0.32\columnwidth]{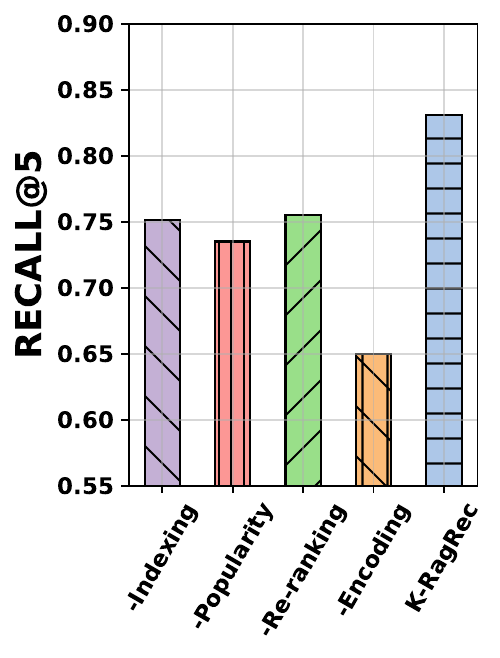}}
\subfigure[BOOK ACC]{\includegraphics[width=0.32\columnwidth]{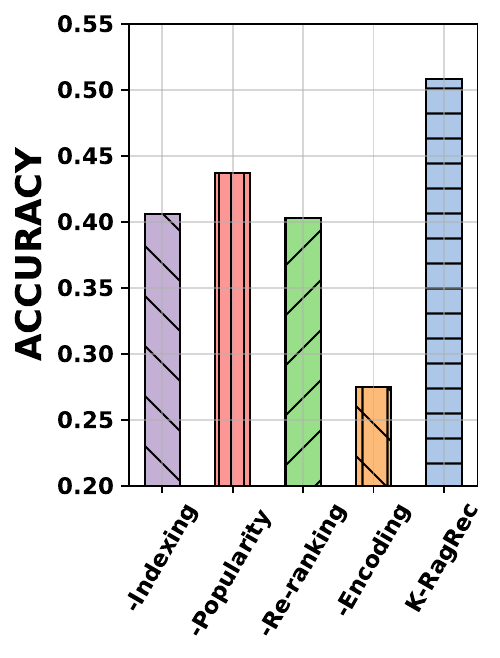}}
\subfigure[BOOK R@3]{\includegraphics[width=0.32\columnwidth]{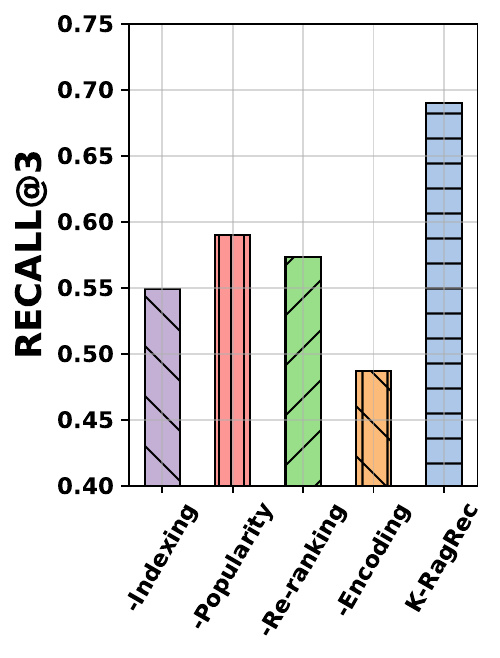}}
\subfigure[BOOK R@5]{\includegraphics[width=0.32\columnwidth]{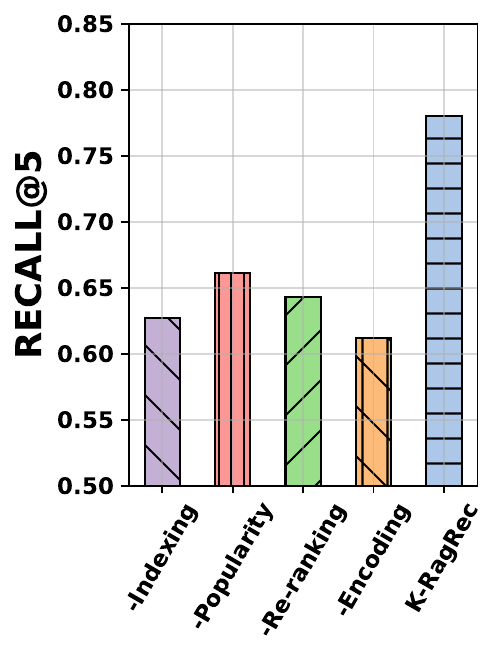}}
\caption{Comparison among K-RagRec and its four ablated variants on MovieLens-1M and Amazon Book datasets and LLama-2-7b across metrics Accuracy, Recall@3 and Recall@5.}\label{fig:ablation}
\end{figure}

\subsection{\textbf{Ablation Study}}
To evaluate the impact of each component in our proposed framework, we conduct the ablation study to compare the K-RagRec with four ablated variants on MovieLens and Amazon Book datasets, using LLama-2-7B as the backbone LLM. Details of each ablation variant are provided in Appendix~\ref{appendix:Abla}. The results are illustrated in Figure~\ref{fig:ablation}.
Observing the experiment results, we find that eliminating any component of the framework leads to a decrease in the overall performance of the recommendations, demonstrating the effectiveness of each module. Secondly, removing the GNN Encoder leads to a 37\% decrease and a 45.9\% decrease in the accuracy of the model on MovieLens and Amazon Book datasets, respectively, highlighting the significance of employing GNN to encode the structure of knowledge sub-graphs. Refer to Appendix~\ref{appendix:Abla} for more details.

\input{tables/inference_time}

\subsection{\textbf{Efficiency Evaluation}}
In this sub-section, we evaluate the inference efficiency of our proposed K-RagRec framework compared with baselines on the MovieLens-1M dataset and LLama-2-7b.
We record the time cost for one inference utilizing two NVIDIA A6000-48G GPUs. The time cost for a single inference is reported in Table~\ref{tab:time}.
By observing the experimental results, we notice that various KG RAG approaches significantly increase the inference time, which is due to the large scale of the KG. In contrast, K-RagRec achieves the best computational efficiency compared to various KG RAG methods and is only about 0.1s slower than direct inference without retrieval.
These findings highlight the efficiency of K-RagRec and validate the effectiveness of our popularity selective retrieval policy.

\subsection{\textbf{Parameter Analysis}}
In this section, we evaluate the impact of three main hyper parameters of K-RagRec, namely popularity selective retrieval policy threshold $p$, retrieved knowledge sub-graph numbers $K$,
and re-ranking knowledge sub-graph numbers $N$.
In addition, we analyze the impact of various GNN encoder variants and different GNN layer numbers for the proposed framework in Appendix~\ref{appendix:gnn encoder} and ~\ref{appendix:gnn layer}.

\textbf{1) Impact of popularity selective retrieval policy threshold $p$}:
To understand how the popularity selective retrieval policy threshold $p$ affects K-RagRec, we conduct experiments on MovieLens-1M and LLama-2-7b across two metrics. Results are shown in Figure~\ref{fig:para1}.
As the threshold $p$ increases, the recommendation performance initially improves and then decreases. When the threshold $p$ is set to a small value, only a few items are augmented. This leads to insufficient retrieval and poor recommendation accuracy. When $p$ is set to a larger value, more items are retrieved for augmentation. However, due to the re-ranking sub-graph numbers $N$ being fixed, some retrieved cold-start item knowledge sub-graphs are discarded or ranked at the back of the list, resulting in sub-optimal recommendation performance. 
On the other hand, as observed in Figure~\ref{fig:para1} (c), the inference time increases almost linearly with threshold $p$. Therefore, selecting an appropriate threshold $p$ is crucial to balance the performance and inference time.

\textbf{2) Impact of retrieved knowledge sub-graph numbers $K$ and re-ranking sub-graph numbers $N$}:
In this part, we analyze the impact of two key hyper parameters, which are retrieved knowledge sub-graph numbers $K$ and re-ranking sub-graph numbers $N$. First, to measure the impact of $K$, we perform experiments on the MovieLens dataset and fix $p=50\%, \ N=5$. As shown in Figure~\ref{fig:parak}, some relevant knowledge sub-graphs may be overlooked when $K=1$. On the other hand, larger values of $K$ can introduce irrelevant information. Therefore, we set $K$ equal to 3 in our experiments. Next, we evaluate the effect of $N$ by fixing $p=50\%$ and $ \ K=3$, and present the results in Figure~\ref{fig:paraK}. We observe that setting $N$ to between 5 and 7 results in improved performance on the Amazon Book dataset.
In general, it is important to carefully choose $K$ and $N$ based on the scale of the dataset and the KG.

%% file: tables/main_results.tex
\begin{table*}[t]
\centering
\vskip -0.150in
  \caption{Performance comparison of different KG RAG-enhanced LLM recommendations. 
  The \colorbox{backred!50}{best performance} and the \colorbox{backblue!75}{second-best performance} are marked in red and blue, respectively. ACC and R@$k$ denote
Accuracy and Recall@$k$, respectively.}
  \vskip -0.05in
    \scalebox{0.6}
{
  \begin{adjustbox}{}
    \begin{tabular}{c|c|c|c|c|c|c|c|c|c|c|c}
    \toprule
    \toprule
    \multirow{2}{*}{Models}  & \multicolumn{2}{c|}{\multirow{2}{*}{Methods}} & \multicolumn{3}{c|}{MovieLens-1M}  & \multicolumn{3}{c|}{MovieLens-20M} & \multicolumn{3}{c}{Amazon Book} \\
    \cmidrule{4-12}
    & \multicolumn{2}{c|}{} & \multicolumn{1}{c|}{ACC} & \multicolumn{1}{c|}{R@3} & \multicolumn{1}{c|}{R@5}& \multicolumn{1}{c|}{ACC} & \multicolumn{1}{c|}{R@3} & \multicolumn{1}{c|}{R@5} & \multicolumn{1}{c|}{ACC} & \multicolumn{1}{c|}{R@3} & \multicolumn{1}{c}{R@5} \\
\midrule
    \multirow{10}{*}{\textbf{LLama-2}}&   \multirow{2}{*}{\textbf{Inference-only}} 
    & KG-Text~\citep{wu2023retrieve}& 0.076  & - & -  & 0.052 &  - &  -  &  0.058  &  -  &  -   \\
    & &KAPING~\citep{baek2023knowledge} & 0.079  & -  & - & 0.069 &  - &  -  & 0.063  &   - &   -  \\
    \cmidrule{2-12}
    &\multirow{5}{*}{\textbf{Frozen LLM w/ PT}} &PT w/ KG-Text& 0.078  & 0.191  & 0.308 &0.051  &0.152 &0.250   &0.074   &0.165   &0.245   \\
    & &GraphToken w/ RAG~\citep{perozzi2024let} & 0.268  &  0.421 & 0.466 & 0.186 &0.433 &0.576   &0.326   & 0.515  & 0.624   \\
    & &G-retriever~\citep{he2024g} &  0.274 & 0.532  & 0.650 & 0.342 &0.619  &0.739   &0.275   &0.487   &0.612 \\
     & & K-RagRec & \colorbox{backblue!75}{0.435}  &  \colorbox{backblue!75}{0.725} & 0.831 &0.600  & \colorbox{backblue!75}{0.850} & \colorbox{backblue!75}{0.913}  &  \colorbox{backblue!75}{0.508}  & \colorbox{backblue!75}{0.690}  &  \colorbox{backblue!75}{0.780} \\
    \cmidrule{3-12}
    && \cellcolor{verylightgray} \textbf{Improvement} & \cellcolor{verylightgray}58.6\%  & \cellcolor{verylightgray}33.0\% & \cellcolor{verylightgray}27.8\%  & \cellcolor{verylightgray}75.4\% &  \cellcolor{verylightgray}37.3\% &  \cellcolor{verylightgray}23.5\%  &  \cellcolor{verylightgray}55.8\%  & \cellcolor{verylightgray}34.0\%   &  \cellcolor{verylightgray}25.0 \%  \\
    \cmidrule{2-12}
    &\multirow{3}{*}{\textbf{Fine-tuning}} &Lora w/ KG-Text&    0.402 & 0.718 & \colorbox{backblue!75}{0.833} &\colorbox{backblue!75}{0.609} &0.848   &0.905   &0.446   &0.648  &0.758 \\
     & & Lora w/ K-RagRec &  \colorbox{backred!50} {0.466} & \colorbox{backred!50} {0.770}  & \colorbox{backred!50} {0.863} & \colorbox{backred!50} {0.637}  & \colorbox{backred!50} {0.872} & \colorbox{backred!50} {0.927}  & \colorbox{backred!50} {0.516}   & \colorbox{backred!50} {0.720}   & \colorbox{backred!50} {0.799} \\
     \cmidrule{3-12}
    && \cellcolor{verylightgray} \textbf{Improvement} &  \cellcolor{verylightgray}15.9\%  & \cellcolor{verylightgray}7.2\% & \cellcolor{verylightgray}3.6\%  & \cellcolor{verylightgray}4.5\% &  \cellcolor{verylightgray}2.7\% &  \cellcolor{verylightgray}2.4\%  &  \cellcolor{verylightgray}15.7\%  & \cellcolor{verylightgray}11.1\%   &   \cellcolor{verylightgray}5.4\%    \\
   \midrule
    \multirow{10}{*}{\textbf{LLama-3}}&   \multirow{2}{*}{\textbf{Inference-only}} & KG-Text~\citep{wu2023retrieve}& 0.095  & - & -  & 0.060 &  - &  -  &  0.054  &  -  &  - \\
    & &KAPING~\citep{baek2023knowledge}& 0.084  & - & -  & 0.069 &  - &  -  & 0.062   &  -  &  - \\
    \cmidrule{2-12}
    &\multirow{5}{*}{\textbf{Frozen LLM w/ PT}} &PT w/ KG-Text& 0.134  & 0.294  & 0.433 &0.094  &0.205 &0.296   &0.083   &0.207   &0.314   \\
    & &GraphToken w/ RAG~\citep{perozzi2024let} & 0.355  & 0.622  & 0.737 & 0.473 &0.719 & 0.805  &0.428   &0.567   &0.661    \\
    & &G-retriever~\citep{he2024g} & 0.352  & 0.632  & 0.746 & 0.502 &0.736 & 0.796  &0.417   & 0.584  &0.682 \\
     & & K-RagRec & \colorbox{backblue!75}{0.472}  & \colorbox{backblue!75}{0.704}  & \colorbox{backblue!75}{0.765} &0.634  &\colorbox{backblue!75}{0.779} &\colorbox{backblue!75}{0.818}    & \colorbox{backblue!75}{0.514}  & \colorbox{backblue!75}{0.662}  & \colorbox{backblue!75}{0.723} \\
  \cmidrule{3-12}
    && \cellcolor{verylightgray} \textbf{Improvement} &  \cellcolor{verylightgray}32.9\%  & \cellcolor{verylightgray}11.4\% & \cellcolor{verylightgray}2.5\%  & \cellcolor{verylightgray}26.3\% &  \cellcolor{verylightgray}5.8\% &  \cellcolor{verylightgray}1.6\%  &  \cellcolor{verylightgray}20.0\%  & \cellcolor{verylightgray}13.4\%   &   \cellcolor{verylightgray}6.0\%     \\
    \cmidrule{2-12}
    &\multirow{3}{*}{\textbf{Fine-tuning}} &Lora w/ KG-Text& 0.449  & 0.694  &0.750 &\colorbox{backblue!75}{0.648}  &0.757 &0.790   &0.490   &0.638   & 0.698 \\
     & & Lora w/ K-RagRec &  \colorbox{backred!50} {0.498} & \colorbox{backred!50} {0.712}  & \colorbox{backred!50} {0.771} & \colorbox{backred!50} {0.674}  & \colorbox{backred!50} {0.786} & \colorbox{backred!50} {0.817}  & \colorbox{backred!50} {0.546}   & \colorbox{backred!50} {0.672}   & \colorbox{backred!50} {0.733}  \\
    \cmidrule{3-12}
    && \cellcolor{verylightgray} \textbf{Improvement} &   \cellcolor{verylightgray}10.9\%  & \cellcolor{verylightgray}2.6\% & \cellcolor{verylightgray}2.8\%  & \cellcolor{verylightgray}4.0\% &  \cellcolor{verylightgray}3.8\% &  \cellcolor{verylightgray}3.4\%  &  \cellcolor{verylightgray}11.4\%  & \cellcolor{verylightgray}5.3\%   &   \cellcolor{verylightgray}5.0\%      \\
   \midrule
    \multirow{10}{*}{\textbf{QWEN2}}&   \multirow{2}{*}{\textbf{Inference-only}} & KG-Text~\citep{wu2023retrieve}&0.160   & - & -  & 0.174 &  - &  -  &  0.194  &  -  &  - \\
    & &KAPING~\citep{baek2023knowledge}& 0.196  & - & -  & 0.208 &  - &  -  &  0.220  &  -  &  - \\
    \cmidrule{2-12}
    &\multirow{5}{*}{\textbf{Frozen LLM w/ PT}} &PT w/ KG-Text& 0.190 &  0.371 & 0.499 &0.259  &0.397 &0.494   &0.303   &0.451   &0.553   \\
    & &GraphToken w/ RAG~\citep{perozzi2024let} & 0.259	& 0.487	& 0.608  & 0.370 &0.550 &0.632   & 0.365  &0.568   &  0.658  \\
    & &G-retriever~\citep{he2024g} & 0.304	&0.551	&0.644  &0.389  &0.606 &0.685   &0.355   &0.552   &0.658 \\
     & & K-RagRec &   \colorbox{backblue!75}{0.416}	&\colorbox{backblue!75}{0.712}	&\colorbox{backblue!75}{0.829}  &0.586  &\colorbox{backblue!75}{0.842}  &0.904   & \colorbox{backblue!75}{0.502}  & \colorbox{backblue!75}{0.686} &  \colorbox{backblue!75}{0.767} \\
         \cmidrule{3-12}
    && \cellcolor{verylightgray} \textbf{Improvement} &   \cellcolor{verylightgray}36.8\%  & \cellcolor{verylightgray}29.2\% & \cellcolor{verylightgray}28.4\%  & \cellcolor{verylightgray}50.6\% &  \cellcolor{verylightgray}38.9\% &  \cellcolor{verylightgray}32.0\%  &  \cellcolor{verylightgray}37.5\%  & \cellcolor{verylightgray}20.8\%   &  \cellcolor{verylightgray}16.6 \%      \\
    \cmidrule{2-12}
    &\multirow{3}{*}{\textbf{Fine-tuning}} &Lora w/ KG-Text&   0.400	&0.701	&0.815  &\colorbox{backblue!75}{0.601}  &\colorbox{backblue!75}{0.842} & \colorbox{backblue!75}{0.906}  &0.478   &0.667   &0.751 \\
     & & Lora w/ K-RagRec &   \colorbox{backred!50} {0.466}	& \colorbox{backred!50} {0.763}	&\colorbox{backred!50} {0.860}  & \colorbox{backred!50} {0.631}  & \colorbox{backred!50} {0.868}  & \colorbox{backred!50} {0.928}   & \colorbox{backred!50} {0.510}  & \colorbox{backred!50} {0.704}  & \colorbox{backred!50} {0.780} \\
    \cmidrule{3-12}
    && \cellcolor{verylightgray} \textbf{Improvement} &   \cellcolor{verylightgray}16.5\%  & \cellcolor{verylightgray}8.8\% & \cellcolor{verylightgray}5.5\%  & \cellcolor{verylightgray}5.0\% & \cellcolor{verylightgray}3.1 \% &  \cellcolor{verylightgray}2.4\%  &  \cellcolor{verylightgray}6.7\%  & \cellcolor{verylightgray}5.5\%   &   \cellcolor{verylightgray}3.9\%      \\
    \bottomrule
    \bottomrule
    \end{tabular}%
    \end{adjustbox}
}
  \label{tab:results}%
\end{table*}%

%% file: tables/inference_time.tex
\begin{table}[t]
  \centering
  \caption{ 
   Comparison of the inference efficiency on the MovieLens-1M dataset and LLama-2-7b in seconds (s). ACC denote Accuracy.}
  \vskip -0.05in
    \scalebox{0.72}
{
    \begin{tabular}{c|c|c}
    \toprule
    \toprule
    \multirow{1}{*}{Methods}  & \multirow{1}{*}{ACC} & \multirow{1}{*}{Time (s)}  \\
    \midrule
    w/o RAG & - & 0.92 \\
    KG-Text  & 0.076 & 2.19 \\
    KAPING & 0.079 & 6.47 \\
    GraphToken w/ RAG & 0.268 & 3.14 \\
    G-retriever & 0.274 & 5.86 \\
    \midrule
    K-RagRec & 0.429 & 1.06 \\
    \bottomrule
    \bottomrule
    \end{tabular}%
}
  \label{tab:time}%
  \vskip -0.18in
\end{table}%

%% file: sections/conclusion.tex
\section{Conclusion}
In this paper, we propose a novel framework \textbf{K-RagRec} to augment the recommendation capability of LLMs by retrieving reliable and up-to-date knowledge from KGs. Specifically, 
we first introduce a GNN and PLM to perform semantic indexing of KGs, enabling both coarse and fine-grained retrieval for KGs. To further improve retrieval efficiency, we introduce a popularity selective retrieval policy that determines whether an item needs to be retrieved based on its popularity. 
Notably, K-RagRec performs more expressive graph encoding of the retrieved knowledge sub-graphs, facilitating the LLM to effectively leverage the structure information and avoid long context input. Extensive experiments conducted on three real-world datasets demonstrate the effectiveness of our proposed framework.

\section{Limitations}
To the best of our knowledge, this work is a pioneering study in investigating knowledge-graph RAG for LLM-based recommendations.  Therefore, we realise that this work still has the following three main limitations that can be improved in the future research:
\vspace{-0.1in}
 \begin{itemize} [leftmargin=*] 
\setlength{\itemsep}{0pt}
\setlength{\parsep}{0pt}
\setlength{\parskip}{0pt}
\item Firstly, due to the GPU resource constraints, we were only able to evaluate our framework on 7b and 8b models. Therefore, in future work, we plan to extend our method to larger models to fully assess its effectiveness and scalability.
\item Additionally, we only utilize Freebase as the external KG as it is most commonly used for recommendation tasks. 
Thus we also aim to adopt other KGs, such as YAGO, DBpedia, and Wikipedia, to better understand how different knowledge sources impact the performance of the proposed method.
\item Lastly, designing an intelligent selective retrieval policy for LLM-based recommender systems is an important and challenging task.
In this work, we propose to leverage popularity to
determine which items to retrieve to improve retrieval efficiency.
In the future, we will investigate more flexible mechanisms (e.g., reinforcement learning) to dynamically update policies according to changes in user interest. 
\end{itemize}

\section{Acknowledgements}
The research described in this paper has been partly supported by General Research Funds from the Hong Kong Research Grants Council (project no. PolyU 15207322, 15200023, 15206024, and 15224524), internal research funds from The Hong Kong Polytechnic University (project no. P0042693, P0048625, P0051361, P0052406, and P0052986).

%% file: sections/appendix.tex
\newpage

\section{Appendix}

\subsection{Implementation Details}\label{appendix:hyper}
The hyper parameters used for K-RagRec and their corresponding values are shown in Table~\ref{tab:hyper}. The first part provides the general training setting for K-RagRec. The second part presents the details of ($\text{GNN}^{\text{Indexing}}$, $\text{GNN}^{\text{Encoding}}$). Then we list the Lora and acceleration settings. Lastly, we provide the hyper parameters for retrieval, including candidate item number $M$, popularity selective retrieval policy threshold $p$, retrieve knowledge sub-graphs numbers $K$, and re-ranking knowledge sub-graphs numbers $N$. If not specified, we run all methods three times with different random seeds and report the averaged results.

\begin{table}[htbp]
 \vskip -0.02in
  \centering
  \caption{Basic statistics of three datasets and the KG. "Items in KG" indicates the number of items that appeared in both the KG and the dataset. 
  }
   \vskip -0.1in
    \scalebox{0.6}
{
  \begin{adjustbox}{}
    \begin{tabular}{c|c|c|c}
    \toprule
    \toprule
    \multirow{1}{*}{Datasets} & {MovieLens-1M} & {MovieLens-20M} & {Amazon Book} \\
    \midrule
User  & 6,038 & 138,287 & 6,106,019 \\
Item & 3,533 & 20,720 & 1,891,460 \\
Interaction & 575,281 & 9,995,410 & 13,886,788 \\
Items in KG & 3,498 & 20,139 & 91,700 \\
\midrule
Entitys & 250,631 & 1,278,544 & 186,954\\
Relations &  264   & 436    & 16 \\
KG triples & 348,979 & 1,827,361 & 259,861  \\
    \bottomrule
    \bottomrule
    \end{tabular}%
    \end{adjustbox}
}
 \vskip -0.1in
  \label{tab:datasets}%
\end{table}%

\subsection{Graph Neural Networks}\label{appendix:gnn}
GNNs are a critical technique in graph machine learning and are widely employed in various graph tasks. 
By iteratively updating node representations through aggregating information from neighboring nodes, GNNs effectively capture the underlying topology and relational structure of graphs.
Formally, a typical GNN operation can be formulated as follows:
\begin{equation}
\mathbf{x}_j^{(l+1)}=\mathbf{x}_j^{(l)} \oplus \mathrm{AGG}^{(l+1)}\left(\left\{\mathbf{x}_i^{(l)} \mid i \in \mathcal{N}_{x_j}\right\}\right),
\end{equation}
where $\mathbf{x}_j^{(l+1)}$ express node $j$'s feature on the $l$-th layer, and $\mathcal{N}(x_j)$ is the set of neighbours of node $j$. $\mathrm{AGG}$ is a aggregation function to aggregates
neighbors’ features, and $\oplus$ combines neighbors' information with the node itself.

\input{tables/hyper_parameters}

\subsection{More Related Work}\label{appendix:related_work}
The remarkable breakthroughs in LLMs have led to their widespread adoption across various fields, particularly in the recommendations. 
Given powerful reasoning and generalization capabilities, many studies have actively attempted to harness the power of the LLM to enhance recommender systems~\citep{geng2022recommendation,bao2023tallrec,qu2024tokenrec,wei2024llmrec,zhang2023collm}. For example, P5~\citep{geng2022recommendation} proposes an LLM-based recommendation model by unifying pre-training, prompting, and prediction for various recommendation tasks, such as sequential recommendation and rating predictions. Furthermore, Tallrec~\citep{bao2023tallrec} fine-tunes LLM (i.e., LLaMA-7B) to align with recommendation data for sequential recommendations.
To further capture higher-order collaborative knowledge and enhance the model's ability to generalize users and items, TokenRec~\citep{qu2024tokenrec} proposes a masked vector-quantized tokenizer to tokenize users and items in LLM-based recommendations.
Despite their effectiveness, these models often face challenges, such as hallucinations and the lack of up-to-date knowledge.
While fine-tuning can partially mitigate these issues, it is resource-costly and time-consuming due to the massive parameters of LLMs.
To overcome these challenges, we propose a knowledge retrieval augmented recommendation framework that leverages external KGs to provide reliable and up-to-date knowledge instead of costly fine-tuning.

\subsection{Comparison with Existing Methods}
Existing LLM-based recommender systems usually require frequent fine-tuning on specific datasets to address the lack of knowledge and hallucinations, which is time-consuming and costly. To solve these challenges and avoid costly fine-tuning, our work is the first to augment the recommendation performance of LLMs by retrieving structured data (i.e., knowledge graph). Specifically, our approach differs from existing work in the following ways:
\vspace{-0.1in}
 \begin{itemize} [leftmargin=*] 
\setlength{\itemsep}{0pt}
\setlength{\parsep}{0pt}
\setlength{\parskip}{0pt}
\item K-RagRec introduces an indexing GNN to efficiently retrieve structured data to enhance the recommendation capability of LLMs. Although some GraphRAG approaches also introduce GNNs to capture higher-order neighbourhood information, they only apply the last layer of the GNN representation, leading to coarse retrieval~\citep{he2024g, mavromatis2024gnn}. In contrast, our approach leverages the representation of each GNN layer to retrieve nuanced knowledge of both coarse and fine-grained graph structures from KG, achieving a more comprehensive and precise retrieval.
\item In the recommendation domain that pursues inference speed, excessive retrieval time can seriously degrade the user experience resulting in user churn. Although many studies have explored selective retrieval for RAGs~\citep{yan2024corrective,jiang2023active}, it is still an open question to determine whether an item needs to be retrieved and to reduce the retrieval time in the recommendation domain. We propose to use popularity to decide whether an item needs to be retrieved based on power law distribution, which greatly reduces the retrieval time. Table~\ref{tab:time} shows that K-RagRec achieves inference times close to direct inference while maintaining high recommendation accuracy.
\item Typical RAG methods usually incorporate the retrieved content into the prompt as text~\citep{wu2023retrieve,baek2023knowledge}. However, vanilla RAG methods rely on documents and paragraphs often introduce unnecessary noise and even harmful disturbance, which can negatively impact the accuracy and reliability of recommendations. In addition, the structural relationships between entities are overlooked in typical RAG, resulting in the sub-optimal reasoning capability of LLM-based recommender systems. We propose to incorporate the retrieved knowledge subgraphs (Knowledge-GraphRAG) into the query as a graph prompt, which facilitates LLMs to better understand the retrieved knowledge subgraphs and avoids long contexts.
\end{itemize}

\subsection{Comparison with 10 Candidate Items}\label{appendix:10samples}
In this section, we conduct additional experiments to evaluate the effectiveness of K-RagRec with a candidate item number $M=10$. In this setting, we randomly select nine negative samples with the target item. The results are presented in Table~\ref{tab:10samples}. We exclude the results of some backbone LLM models (e.g., LLama-3 and QWEN2), as similar observations as Table~\ref{tab:results} can be found. Observing the experimental results, we can note that our proposed K-RagRec method consistently outperforms all baseline methods on MovieLens and Amazon Book datasets, further highlighting the effectiveness of our framework.

\input{tables/10samples}

\subsection{Ablation Study Setting}\label{appendix:Abla}
To assess the impact of each module in K-RagRec, we compare the framework with four ablated variants: K-RagRec (\textit{-Indexing}), K-RagRec (\textit{-Popularity}), K-RagRec (\textit{-Re-ranking}), and K-RagRec (\textit{-Encoding}). (1) K-RagRec (\textit{-Indexing}) eliminates the $\text{GNN}^{\text{Indexing}}$ and stores semantic information of PLM in the knowledge vector database. For the retrieved nodes, we extract their second-order sub-graphs as the retrieved knowledge sub-graphs. (2) K-RagRec (\textit{-Popularity}) does not apply the popularity selective retrieval policy and retrieves all items from the user's historical interactions. (3) (\textit{-Re-ranking}) removes the re-ranking module and inputs the knowledge sub-graphs directly. (4) K-RagRec (\textit{-Encoding}) removes the $\text{GNN}^{\text{Encoding}}$ and replaces it with a trainable soft prompt. The retrieved knowledge sub-graphs will be added to the prompt as triples (e.g., \textit{\{Moonraker, film writer film, Christopher Wood (writer)\}}).

\subsection{\textbf{Generalization Study}}\label{appendix:General}
To evaluate the generalization capability of our proposed framework in the zero-shot setting, we trained a version of the model on the MovieLens-1M dataset and assessed it on the MovieLens-20M and Amazon Book datasets. The experiment results are shown in Table~\ref{tab:transferability}. 
We note that although the K-RagRec performance in the zero-shot setting is slightly degraded when compared to the well-trained model in Table~\ref{tab:results},  it still demonstrates 21.6\% improvement over SOTA baselines on the MovieLens-20M dataset. 
Furthermore, despite the differences between the book recommendation and movie recommendation tasks, the model trained on MovieLens-1M delivers about 8.7\% improvement in the zero-shot setting compared to prompt-tuned baselines on the Amazon Book dataset.
The experimental results demonstrate that K-RagRec exhibits strong generalization capabilities and is adaptable across different domains.

\input{tables/transfer}

\input{tables/hallucination}

\subsection{Study of Hallucination}
In this section, we present a qualitative analysis of hallucinations in the LLama-2-7b and QWEN2 models on the MovieLens dataset. Specifically, we include a few fictional movies in the candidate items to observe the probability of the fictional movie being recommended. We compare direct recommendations and recommendations augmented with K-RagRec, and the results are shown in Table~\ref{tab:hallucination}. We note that K-RagRec significantly reduced hallucinations by 93.1\% compared to direct inference on LLama-2. In contrast to LLama-2, QWEN2 rarely recommends fictional movies. Nevertheless, K-RagRec reduced hallucinations by 80.9\%, demonstrating the effectiveness of our approach in addressing hallucinations.

\input{tables/cold_start}

\subsection{Study of Cold Start Recommendation}
The cold start problem is an important issue in most recommendation research. To comprehensively evaluate our approach, we particularly design a case study to evaluate the model's recommendation performance under the cold-start setting. Specifically, we construct a separate cold-start dataset based on the MovieLens dataset that only contains these identified cold-start items as target items. We compare K-RagRec with three KG RAG-enhanced LLM recommendation methods, and the experiment results are shown in Table~\ref{tab:cold_start}. The results demonstrate that our proposed K-RagRec still has satisfactory performance under the cold-start recommendation scenario, highlighting the effectiveness of our framework in all the cases.

\input{tables/gnn}
\subsection{Study of Four GNN Encoders}\label{appendix:gnn encoder}
To further understand the generality of our proposed approach, we conduct a comparative study of four variants applying different GNN encoders. Specifically, we compare GCN~\citep{kipf2016semi}, GAT~\citep{velickovic2017graph}, Graph Transformer~\citep{shi2020masked}, and GraphSAGE~\citep{hamilton2017inductive} as four K-RagRec variants of GNN encoder.
Specifically, Graph Convolutional Network (GCN)~\citep{kipf2016semi} first introduces convolutional operations to graph-structured data. By aggregating features from neighboring nodes, GCN facilitates the learning of rich node representations. GraphSAGE~\citep{hamilton2017inductive} learns an aggregation function that samples and combines features from a node's local neighborhood in an inductive setting, enabling the effective use of new nodes.
Graph Attention Network (GAT)~\citep{velickovic2017graph} further incorporates attention mechanisms, allowing the model to dynamically assign varying attention to neighboring nodes, thereby enhancing the focus on the most relevant information.
Inspired by the success of the transformer, the Graph Transformer~\citep{shi2020masked} adapts transformer architectures to graph data, enhancing the modeling of graphs, particularly textual graphs.

We report the experiment results on the MovieLens-1M dataset and the LLama-2-7b backbone in Table~\ref{tab:gnn}. 
It is noted that the GCN Encoder variant method performs second-best on the Recall@5 metric, although it is slightly worse than other GNN encoders on the Accuracy metric.
Overall, four GNN encoder variants exhibit close performance on the MovieLens dataset, highlighting the generality and robustness of our framework across different GNN encoders.

\input{tables/gnn_layers}
\subsection{Study of GNN Layer Numbers}\label{appendix:gnn layer}
In this subsection, we evaluate the impact of the number of GNN layers on model performance. We vary the numbers of the GNN layer numbers in the range of \{3, 4, 5\} and test on the Amazon Book dataset and LLama-2-7b across three metrics. As observed in Table~\ref{tab:gnn_layer}, the model performance first improves and then decreases as the number of GNN layers increases, and the model achieves the best results across the three metrics when setting the number of GNN layers is set to four. 
Therefore, a smaller number of GNN layers may not have sufficient depth to capture the intricate relationships and dependencies in the graph, leading to sub-optimal performance. On the other hand, too many layers can result in indistinguishable node representations. Thus, selecting the optimal number of GNN layers is crucial for effective model training.

\begin{figure*}[htbp]
\centering
\subfigure[Accuracy]{\includegraphics[width=0.65\columnwidth]{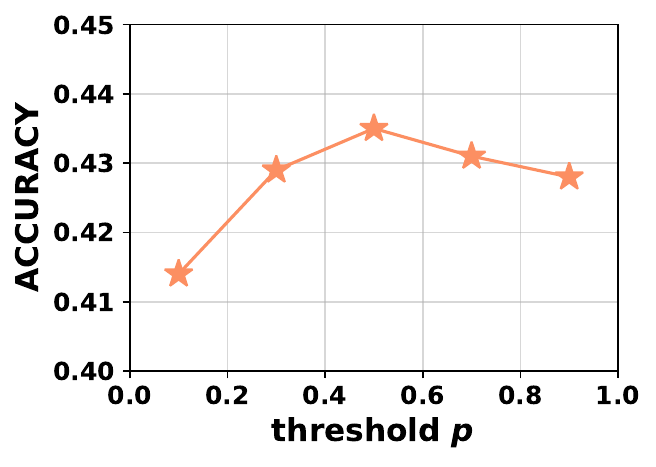}}
\subfigure[Recall@5]{\includegraphics[width=0.65\columnwidth]{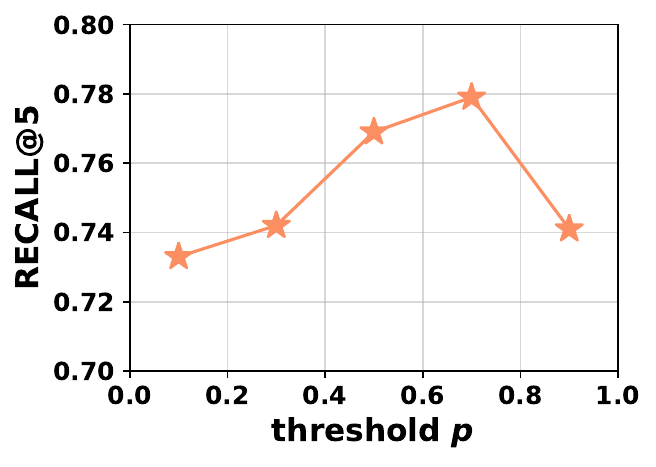}}
\subfigure[Inference Time]{\includegraphics[width=0.65\columnwidth]{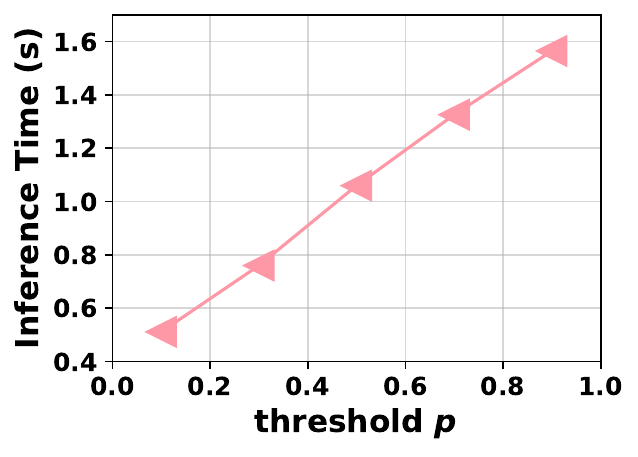}}

\vskip -0.13in
\caption{
Effect of popularity selective retrieval policy threshold $p$ on MovieLens-1M and LLama-2-7b across metrics Accuracy,
Recall@5 and inference time (seconds) for K-RagRec.
}\label{fig:para1}
\vskip -0.23in
\end{figure*}

\begin{figure*}[htbp]
\centering
\subfigure[Amazon Book: Accuracy]{\includegraphics[width=0.65\columnwidth]{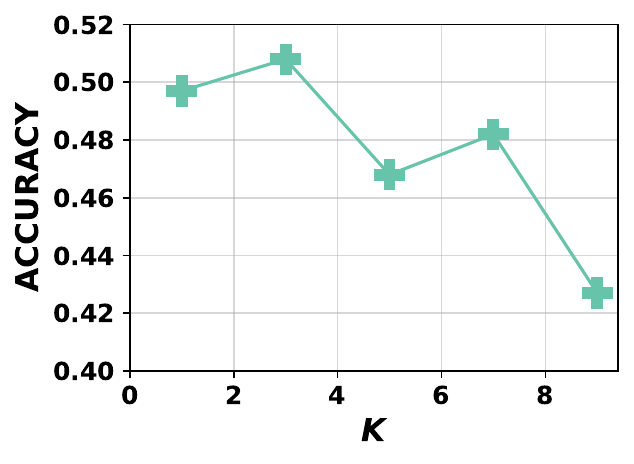}}
\subfigure[Amazon Book: Recall@3]{\includegraphics[width=0.65\columnwidth]{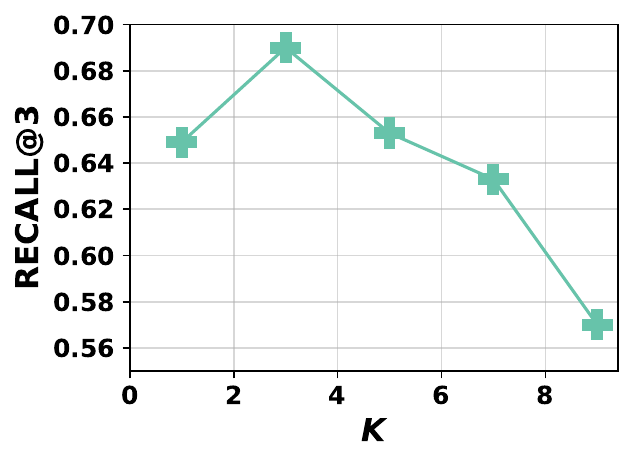}}
\subfigure[Amazon Book: Recall@5]{\includegraphics[width=0.65\columnwidth]{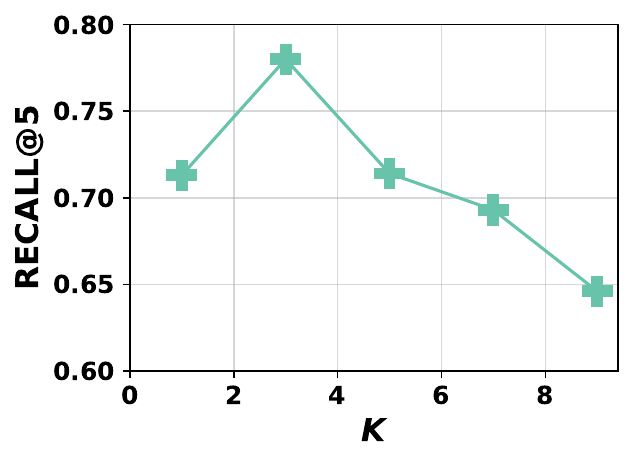}}

\vskip -0.13in
\caption{
Effect of retrieved knowledge sub-graph numbers $K$ on Amazon Book datasets and LLama-2-7b across metrics Accuracy,
Recall@3 and Recall@5 for K-RagRec.
}\label{fig:parak}
\vskip -0.23in
\end{figure*}

\begin{figure*}[htbp]
\centering
\subfigure[Amazon Book: Accuracy]{\includegraphics[width=0.65\columnwidth]{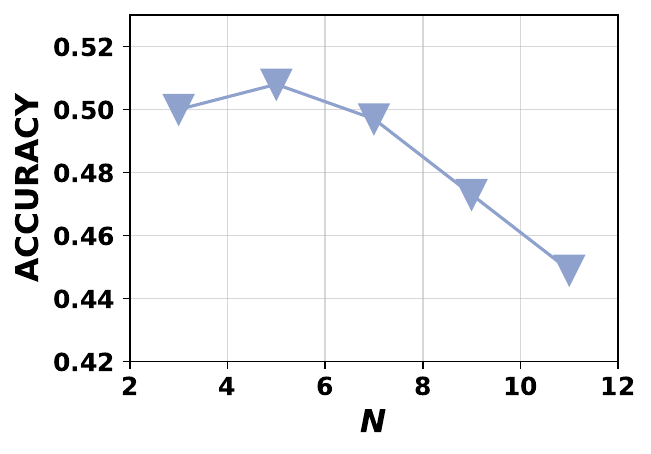}}
\subfigure[Amazon Book: Recall@3]{\includegraphics[width=0.65\columnwidth]{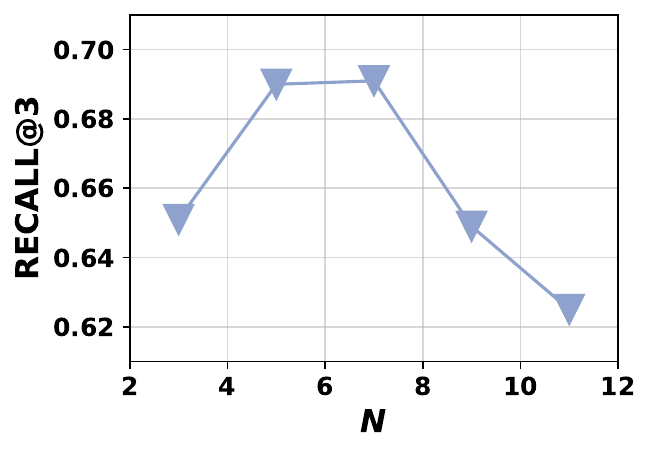}}
\subfigure[Amazon Book: Recall@5]{\includegraphics[width=0.65\columnwidth]{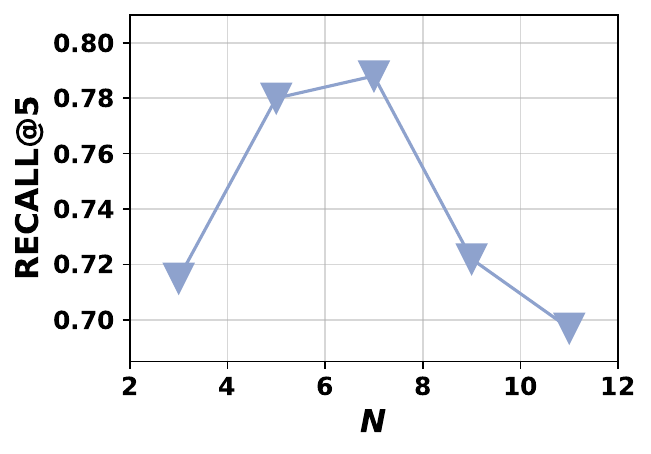}}

\vskip -0.13in
\caption{
Effect of re-ranking knowledge sub-graph numbers $N$ on Amazon Book datasets and LLama-2-7b across metrics Accuracy,
Recall@3 and Recall@5 for K-RagRec.
}\label{fig:paraK}
\vskip -0.13in
\end{figure*}

\subsection{Used Prompt}\label{appendix:prompt}
In this part, we present the prompts designed for movie recommendations and book recommendations. We show two examples in Table~\ref{tab:prompt}, and set the candidate items $M$ equal to 20. For inference, we leverage the model to make the prediction based on the user's recent watching history and candidate items.

\input{tables/prompt}

%% file: tables/hyper_parameters.tex
\begin{table}[htbp]
  \centering
  \caption{Statistics of Hyper Parameters.}
    \scalebox{0.8}
{
    \begin{tabular}{c|c}
    \toprule
    \toprule
    \multirow{1}{*}{Item}  & \multirow{1}{*}{Value}  \\
    \midrule
    batch size & 5  \\
    epochs  & 3  \\
    grad steps & 2 \\
    learning rate & 1e-5 \\
    \midrule
    Indexing layer numbers & 4  \\
    Indexing hidden dimension & 1024  \\
    Encoding layer numbers & 4  \\
    Encoding hidden dimension & 1024  \\
    Encoding head numbers & 4  \\
    \midrule
    lora\_r & 8 \\
    lora\_alpha & 16 \\
    lora\_dropout & 0.1 \\
    int8 & True \\
    fp16 & True \\
    \midrule
    candidate item numbers & 20\\
    threshold $p$ & 50\% \\
    top-$K$ & 3 \\
    top-$N$ & 5 \\
    \bottomrule
    \bottomrule
    \end{tabular}%
}
  \label{tab:hyper}%
\end{table}%

%% file: tables/10samples.tex
\begin{table}[t]
  \centering
  \caption{ Performance comparison of different KG RAG-enhanced LLM recommendations with candidate item numbers $M=10$ on the MovieLens and Amazon Book dataset and LLama-2-7b across
two metrics. The best performances are labeled in bold. ACC and R@3 denote Accuracy and Recall@3, respectively. }
    \scalebox{0.65}
{
    \begin{tabular}{c|c|c|c|c}
    \toprule
    \toprule
    \multirow{2}{*}{Methods} & \multicolumn{2}{c|}{MovieLens-1M} & \multicolumn{2}{c}{Amazon Book} \\
    \cmidrule{2-5}
    & \multicolumn{1}{c|}{ACC} & \multicolumn{1}{c|}{R@3} & \multicolumn{1}{c|}{ACC} & \multicolumn{1}{c}{R@3}  \\
    \midrule
    KG-Text &0.185  &- &0.142  &-  \\
    KAPING &0.165  &- & 0.119 &-  \\
    PT w/ KG-Text &0.159  &0.493 &0.123  &0.384  \\
    GraphToken w/ RAG &0.512  &0.753 &0.444  & 0.682 \\
    G-retriever &0.469  &0.721 &0.367  &0.610  \\
    \midrule
 K-RagRec &\textbf{0.568}  &\textbf{0.779} &\textbf{0.606}  &\textbf{0.770}  \\
    \bottomrule
    \bottomrule
    \end{tabular}%
}
  \label{tab:10samples}%
\end{table}%

%% file: tables/transfer.tex
\begin{table}[t]
  \centering
  \caption{ The generalization results for our K-RagRec model in a zero-shot setting. In this setting, our models are trained on MovieLens-1M dataset and evaluated on MovieLens-20M and Amazon Book datasets.  ACC and R@$k$ denote Accuracy
and Recall@$k$, respectively.}
  \vskip -0.05in
    \scalebox{0.57}
{
    \begin{tabular}{c|c|c|c|c|c|c|c}
    \toprule
    \toprule
    \multirow{2}{*}{Models}  & \multirow{2}{*}{Methods} & \multicolumn{3}{c|}{MovieLens-20M} & \multicolumn{3}{c}{Amazon Book} \\
    \cmidrule{3-8}
    & & \multicolumn{1}{c|}{ACC} & \multicolumn{1}{c|}{R@3} & \multicolumn{1}{c|}{R@5} & \multicolumn{1}{c|}{ACC} & \multicolumn{1}{c|}{R@3} & \multicolumn{1}{c}{R@5} \\
    \midrule
 \multirow{2}{*}{LLama2} & K-RagRec &0.539  &0.740 &0.795 &0.390  &0.581 &0.671    \\
  & Lora w/K-RagRec &0.539  &0.783 &0.863 &0.405 &0.580 &0.796   \\
  \midrule
\multirow{2}{*}{LLama3} &K-RagRec  &0.597 &0.797 &0.839 &0.428 &0.628 &0.706 \\
 & Lora w/K-RagRec  &0.611  &0.775 &0.814 &0.424 &0.622 &0.732  \\
 \midrule
\multirow{2}{*}{QWEN} &K-RagRec  &0.507 &0.769 &0.861 &0.418 &0.612  &0.687\\
 & Lora w/K-RagRec &0.545  &0.814 &0.897 &0.441 &0.623 &0.706   \\
    \bottomrule
    \bottomrule
    \end{tabular}%
}
  \label{tab:transferability}%
  \vskip -0.06in
\end{table}%

%% file: tables/hallucination.tex
\begin{table}[t]
  \centering
  \caption{Quantitative comparison of hallucination on the MovieLens-1M dataset. $\Delta$ denotes the reduction in hallucinations for K-RagRec compared to Direct Inference.}
    \scalebox{0.75}
{
    \begin{tabular}{c|c|c|c}
    \toprule
    \toprule
    \multirow{1}{*}{Models}
    & \multicolumn{1}{c|}{Direct Inference} & \multicolumn{1}{c|}{K-RagRec} & \multicolumn{1}{c}{$\Delta$}\\
    \midrule
    LLama-2 &39.1\%  &2.7\% &93.1\% \\
    \midrule
 QWEN2 &4.7\% &0.9\% & 80.9\% \\
    \bottomrule
    \bottomrule
    \end{tabular}%
}
  \label{tab:hallucination}%
\end{table}%

%% file: tables/cold_start.tex
\begin{table}[t]
  \centering
  \caption{ Performance comparison of different KG RAG-enhanced LLM recommendation methods on the cold-start dataset and QWEN2 across three metrics. The best performances are labeled in bold. ACC and R@$k$ denote Accuracy and Recall@$k$, respectively. }
    \scalebox{0.85}
{
    \begin{tabular}{c|c|c|c}
    \toprule
    \toprule
    \multirow{1}{*}{Methods}
    & \multicolumn{1}{c|}{ACC} & \multicolumn{1}{c|}{R@3} & \multicolumn{1}{c}{R@5} \\
    \midrule
    PT w/ KG-Text &0.106  &0.239 &0.395  \\
    GraphToken w/ RAG &0.258  &0.473 &0.620 \\
    G-retriever &0.185  &0.384 &0.488  \\
    \midrule
 K-RagRec &\textbf{0.406}  &\textbf{0.705} &\textbf{0.834}  \\
    \bottomrule
    \bottomrule
    \end{tabular}%
}
  \label{tab:cold_start}%
\end{table}%

%% file: tables/gnn.tex
\begin{table}[t]
  \centering
  \caption{ 
   Comparison of different GNN Encoders on the MovieLens-1M dataset and LLama-2-7b across three metrics. We use bold fonts to label the best performance. ACC and R@k denote Accuracy and Recall@k, respectively.}
  \vskip -0.05in
    \scalebox{0.75}
{
    \begin{tabular}{c|c|c|c}
    \toprule
    \toprule
    \multirow{1}{*}{GNN Types}  & \multirow{1}{*}{ACC} & \multirow{1}{*}{R@3} &\multirow{1}{*}{R@5}  \\
    \midrule
    GCN~\citep{kipf2016semi} & 0.397 & 0.704 & 0.809 \\
    GAT~\citep{velickovic2017graph}  & 0.420 & 0.693 & 0.804 \\
    Graph Transformer~\citep{shi2020masked} & \textbf{0.429} & \textbf{0.711} &0.779 \\
    GraphSAGE~\citep{hamilton2017inductive} &0.418  &0.699  &\textbf{0.823} \\
    \bottomrule
    \bottomrule
    \end{tabular}%
}
  \label{tab:gnn}%
  \vskip -0.06in
\end{table}%

%% file: tables/gnn_layers.tex
\begin{table}[t]
  \centering
  \caption{ 
   Comparison of different GNN layers on the Amazon Book dataset and LLama-2-7b across three metrics. ACC and R@k denote Accuracy and Recall@k, respectively.}
    \scalebox{0.75}
{
    \begin{tabular}{c|c|c|c}
    \toprule
    \toprule
    \multirow{1}{*}{GNN Layers}  & \multirow{1}{*}{ACC} & \multirow{1}{*}{R@3} &\multirow{1}{*}{R@5}  \\
    \midrule
    3 layers&  0.496 & 0.653 & 0.736 \\
    4 layers& 0.506 & 0.690 & 0.780 \\
    5 layers& 0.498 &0.656 &0.729 \\
    \bottomrule
    \bottomrule
    \end{tabular}%
}
  \label{tab:gnn_layer}%
\end{table}%

%% file: tables/prompt.tex
\begin{table*}[t]
  \centering
  \caption{Example of the used prompt for K-RagRec. The user's recent watching/reading history and candidate items are marked in red and blue, respectively.}
    \scalebox{0.8}{
  \begin{tabularx}{\textwidth}{l|X}
    \toprule
    \multicolumn{1}{c|}{Datasets}  & \multicolumn{1}{c}{Used Prompt} \\
    \midrule
       Movies & Below is an instruction that describes a task, paired with an input that provides further context. Write a response that appropriately completes the request. Instruction: Given the user's watching history, select a film that is most likely to interest the user from the options.  Watching history: \red{\{"History of the World: Part I", "Romancing the Stone", "Fast Times at Ridgemont High", "Good Morning, Vietnam", "Working Girl", "Cocoon", "Splash", "Pretty in Pink", "Terms of Endearment", "Bull Durham"\}}. Options: \blue{\{A: "Whole Nine Yards", B: "Hearts and Minds", C: "League of Their Own", D: "Raising Arizona", E: "Happy Gilmore", F: "Brokedown Palace", G: "Man Who Knew Too Much", H: "Light of Day", I: "Tin Drum", J: "Blair Witch Project", K: "Red Sorghum", L: "Flintstones in Viva Rock Vegas", M: "Anna, N: Roger \& Me" O: "Land and Freedom", P: "In Love and War", Q: "Go West", R: "Kazaam", S: "Thieves", T: "Friends \& Lovers"\}}. Select a movie from options A to T that the user is most likely to be interested in. \\
       \midrule
       Books & Below is an instruction that describes a task, paired with an input that provides further context. Write a response that appropriately completes the request. Instruction: Given the user's reading history, select a book that is most likely to interest the user from the options.  Watching history: \red{\{"Practice Makes Perfect: Spanish Verb Tenses", "NTC's Dictionary of Common Mistakes in Spanish", "Streetwise Spanish Dictionary/Thesaurus", "Buscalo! (Look It Up!) : A Quick Reference Guide to Spanish Grammar and Usage", "La lengua que heredamos: Curso de espaol para bilinges", "Vox Diccionario De Sinonimos Y Antonimos", "Schaum's Outline of Spanish Vocabulary", "Nos Comunicamos (Spanish Edition)", "The Oxford Spanish Business Dictionary", "Bilingual Dictionary of Latin American Spanish"\}}. Options: \blue{\{"A: Folk and Fairy Tales, Childcraft (Volume 3)", B: "The Waves", C: "Dead End Kids: Gang Girls and the Boys They Know", D: "Opera Stars in the Sun: Intimate Glimpes of Metropolitan Opera Personalities", E: "Five-Minute Erotica", F: "Spanish Verbs: Oxford Minireference", G: "Motorcycle Maintenance Techbook: Servicing \& Minor Repairs for All Motorcycles \& Scooters", H: "Father and Son: A Study of Two Temperaments (Classic, 20th-Century, Penguin)", I: "Manga Mania Fantasy Worlds: How to Draw the Amazing Worlds of Japanese Comics", J: "MCSE Designing a Windows Server 2003 Active Directory \& Network Infrastructure: Exam 70-297 Study Guide and DVD Training System", K: "The Atmospheric Boundary Layer (Cambridge Atmospheric and Space Science Series)", L: "St. Augustine and St. Johns County: A pictorial history", M: "A Will to Survive: Indigenous Essays on the Politics of Culture, Language, and Identity", N: "Saved: A Guide to Success With Your Shelter Dog", O: "Mosaic (Star Trek Voyager)", P: "Fantastical Tarot: 78-Card Deck", Q: "American Sign Language-A Look at Its History, Structure and Community", R: "Warrior's Heart (Zebra Historical Romance)", S: "The New Money Management: A Framework for Asset Allocation", T: "Megabrain"\}}. Select a book from options A to T that the user is most likely to be interested in. \\
    \bottomrule
    \end{tabularx}%
    }
  \label{tab:prompt}%
\end{table*}%